\documentclass[sn-aps,Numbered]{sn-jnl}% Math and Physical Sciences

\usepackage{graphicx}%
\usepackage{multirow}%
\usepackage{amsmath,amssymb,amsfonts}%
\usepackage{amsthm}%
\usepackage{mathrsfs}%
\usepackage[title]{appendix}%
\usepackage{xcolor}%
\usepackage{textcomp}%
\usepackage{manyfoot}%
\usepackage{booktabs}%
\usepackage{algorithm}%
\usepackage{algorithmicx}%
\usepackage{algpseudocode}%
\usepackage{listings}%
\usepackage{natbib}
\usepackage{siunitx}%

\usepackage[english]{babel}
\usepackage{dcolumn}% Align table columns on decimal point
\usepackage[utf8]{inputenc} % Required for inputting international characters
\usepackage{bm}% bold math
\usepackage{hyperref}% add hypertext capabilities
\usepackage{mathrsfs}
\usepackage{longtable}
\usepackage{multirow,booktabs,dcolumn,tabularx}
\usepackage{float}

\usepackage{braket}
\usepackage{microtype} % Better typography
\usepackage[dvipsnames]{xcolor}

\usepackage{amsmath,amsfonts,amsthm} % Math packages for equations


\usepackage{dsfont}
\usepackage{comment}
\usepackage{physics}
\usepackage{booktabs}
\usepackage{pifont}
\usepackage{makecell}

\definecolor{limegreen}{rgb}{0.2, 0.8, 0.2}
\definecolor{orange}{rgb}{1.0, 0.5, 0.0}

\definecolor{emerald}{rgb}{0.31, 0.78, 0.47}
\definecolor{blue(ncs)}{rgb}{0.0, 0.53, 0.74}

\begin{document}

\title{Scanning tunneling spectroscopy of superconducting nitridized aluminum thin films}

\author[1]{\fnm{Jose Antonio} \sur{Moreno}}
\author[1]{\fnm{Pablo} \sur{Garc\'ia Talavera}}
\author[2,3]{\fnm{Alba} \sur{Torras-Coloma}}
\author[4]{\fnm{Gemma} \sur{Rius}}
\author[2,5]{\fnm{P.} \sur{Forn-D\'iaz}}
\author[1]{\fnm{Edwin} \sur{Herrera Vasco}}
\author[1]{\fnm{Isabel} \sur{Guillam\'on}}
\author*[1]{\fnm{Hermann} \sur{Suderow}}\email{hermann.suderow@uam.es}

\affil[1]{\orgdiv{Laboratorio de Bajas Temperaturas y Altos Campos Magn\'eticos, Departamento de F\'isica de la Materia Condensada, Instituto Nicol\'as Cabrera and Condensed Matter Physics Center (IFIMAC), Unidad Asociada UAM-CSIC}, \orgname{Universidad Aut\'onoma de Madrid}, \orgaddress{ \city{Madrid}, \postcode{E-28049}, \country{Spain}}}

\affil[2]{\orgdiv{Institut de F\'isica d’Altes Energies (IFAE)}, \orgname{The Barcelona Institute of Science and Technology (BIST)}, \orgaddress{ \city{Bellaterra}, \postcode{E-08193}, \country{Spain}}}

\affil[3]{\orgdiv{Departament de F\'isica}, \orgname{Universitat Aut\`onoma de Barcelona}, \orgaddress{ \city{Bellaterra}, \postcode{E-08193}, \country{Spain}}}

\affil[4]{\orgdiv{Institute of Microelectronics of Barcelona (IMB-CNM)}, \orgname{Consejo Superior de Investigaciones Científicas (CSIC)}, \orgaddress{ \city{Cerdanyola}, \postcode{E-08193}, \country{Spain}}}

\affil[5]{\orgname{Qilimanjaro Quantum Tech SL}, \orgaddress{ \city{Barcelona}, \postcode{E-08007}, \country{Spain}}}

\abstract{
Nitride-based superconductors represent a family of superconducting thin film materials displaying higher quality than their corresponding bare superconductor when used in devices for applications such as cosmic radiation sensing. In recent times, Niobium-based and Titanium-based nitrides were used to improve the quality of superconducting devices in quantum technology applications. Recently, nitridized Aluminum (NitrAl) has been found to display higher critical temperatures and enhanced resilience to magnetic fields compared to those of Al, making it a new interesting candidate for superconducting quantum circuit applications. However, the microscopic properties of NitrAl remain highly unexplored. Here we use Scanning Tunneling Microscope (STM) to measure the superconducting density of states of a thin film sample of nitridized-Aluminum (NitrAl), with a room temperature resistivity between pure Al and fully insulating aluminum nitride. We show that the in-gap density of states is zero up to about $\hbar\omega=250~\mathrm{\mu eV}$ and that there is a distribution of values of the superconducting gap around $\Delta_0=360~\mathrm{\mu eV}$, close to the BCS expectation $\Delta=1.76 k_{\mathrm{B}}T_{\mathrm{c}}$. We also find varying superconducting gap values at the nanometer scale, by approximately 10\%, when probing different regions of the sample. These results suggest a gap which is larger than the one of pure Al, and is spatially more homogeneous than the superconducting gap values often found in thin films. Our work demonstrates that STM is as a powerful tool to screen materials for quantum devices through the measurement of the spatial dependence of the superconducting density of states.}

    
\maketitle
\section*{Introduction}

In an ideal BCS superconductor, the energy gap $\Delta$ produces a vanishing density of states (DOS) at the Fermi level for energies below $ |\Delta|$, while the equilibrium quasiparticle population is exponentially suppressed at low temperatures ($\propto e^{-\Delta/k_B \,T}$)\,\cite{tinkham}. Aluminum is often the preferred choice in superconducting quantum circuit fabrication due to its simple deposition, integration with standard lithography processes and compatibility with Al/AlOx/Al Josephson junction fabrication\,\cite{kim2025josephsonjunctionsagequantum,doi:10.1126/sciadv.abc5055,10.1063/PT.3.5291,Riste2013,Google2019,Google2021,Google2025}. For temperatures below 100 mK, quasiparticles are expected to be absent in Aluminum-based Josephson junctions. However, experiments in the millikelvin range have shown that tunneling of quasiparticles is responsible for the reduction of superconducting qubit lifetimes\,\cite{Aumentado2004,Schreier2008,Martinis2009}.

In addition, multiple relaxation-based effects exist besides quasiparticles, such as two-level systems from surface defects, charge noise and $1/f$ noise. Although there are multiple sources of decoherence, it is thought that engineering the properties of superconducting  materials can be useful to address decoherence\,\cite{Catelani2022}. A recent example is the synthesis and study of Granular Aluminum (GrAl) thin films\,\cite{Grunhaupt2019,Rotzinger2017,Torrascoloma2025}. GrAl consists of superconducting aluminum grains randomly distributed within an insulating oxide matrix\,\cite{Deutscher1973}. GrAl presents an enhancement of the critical temperature $T_{\mathrm{c}}$ and other advantages with respect to Al depending on the growth conditions and composition\,\cite{Abeles1966,Cohen1968,Grunhaupt2018,LevyB2019,Pracht2016}. However, GrAl is based on imperfect oxides, i.e., one of the possible sources of two level systems. One possible route to avoid these oxides is the development of superconducting nitride thin films.

Nitride superconductors were developed for particle detection as kinetic inductance detectors. Recently, these were also considered for integration in quantum circuits\,\cite{Barends2010,Vissers2010,Niepce2019,Leduc2010,Annunziata2010}. The recent discovery of superconducting nitridized aluminum (NitrAl)\,\cite{TorrasColoma2024} shows great promise as a new platform to develop superconducting quantum circuits. NitrAl presents enhanced $T_{\mathrm{c}}$ as compared to pure Al and remains superconducting for magnetic fields well above those of pure Al\,\cite{TorrasColoma2024,Postolova2020,Cordoba2013}. However, microscopic properties of NitrAl, such as the superconducting density of states, have not yet been characterized.

Here, we performed measurements of the superconducting density of states using very low temperature (0.1$~\mathrm{K}$) Scanning Tunneling Microscopy (STM) on a $100~\mathrm{nm}$-thick NitrAl thin film featuring a $T_{\mathrm{c}}=2.4~\mathrm{K}$ and electrical resistivity at 4~K $\rho_{\mathrm 4K} = 48.5~\mu\Omega\,\mathrm{cm}$\cite{TorrasColoma2024}. Unlike GrAl, we find that the topography of this NitrAl thin film shows a smooth surface. In addition, we have found a metallic behavior and a fully developed superconducting gap with $\sim0.37~\mathrm{mV}$, close to the BCS value $\Delta=1.76k_\mathrm{B}T_\mathrm{c}$ for $T_\mathrm{c}\sim2.4~\mathrm{K}$. We have also found superconductivity up to a magnetic field of $\sim500~\mathrm{mT}$ applied out-of-plane. We discuss the implications of the shape of the energy gap and local variations of the superconducting density of states for the use of NitrAl in superconducting quantum circuits.

\section*{Experimental Methods}
The NitrAl thin film was grown by sputtering deposition on a silicon chip at the clean room of the IMB-CNM-CSIC using a high purity $99.99 \%$ Al target and $>5.5~N$ $N_2$ and Ar\,\cite{TorrasColoma2024}. Further details can be found in Ref.\,\cite{TorrasColoma2024}. The square resistance of the film is nearly independent of temperature. The sample was grown with a $N_2$/Ar flow of 8.33~\% which corresponds to sample ``G'' in Ref.\,\cite{TorrasColoma2024}. The superconducting critical temperature is of T$_c=2.4~\mathrm{K}$ and the transition width is below 40$~\mathrm{mK}$\,\cite{TorrasColoma2024}. We have measured the thin film using a home-made STM that is thermally fixed to a dilution refrigerator of Oxford Instruments and cools down to 100$~\mathrm{mK}$. The sample was glued onto a STM sample holder using conductive silver epoxy. A silver epoxy path connected the sample holder with the NitrAl film. We used a Pt-Ir tip which was cleaned and prepared in situ following Ref.\,\cite{Rodrigo2004}. Our STM setup, which includes home-made ultra low noise STM electronics, is described in Refs.\,\cite{Suderow2011, Fernandez2021, Fran2021,githublbtuam}. The setup allows for an energy resolution of $~8~\mu\mathrm{eV}$\,\cite{Fernandez2021}.  Magnetic fields were applied using a superconducting coil from Oxford Instruments perpendicular to the surface of the film. Cooling and other cryogenic arrangements are described in Ref.\,\cite{MONTOYA2019e00058}. For image treatment we use software described in Ref.\,\cite{Fran2021} and available in Ref.\,\cite{githublbtuam}, as well as usual STM software\,\cite{horcas07}.

Maps of the tunneling conductance are obtained by cutting the feedback loop at each point of a topography image and ramping the bias voltage to obtain a current vs voltage curve. Subsequently, we take the derivative of the curve to obtain the tunneling conductance as a function of bias voltage\,\cite{Fernandez2021}. We acquire enough points in the current vs voltage curve to guarantee the resolution in energy discussed before. We usually normalize the conductance vs bias voltage at a voltage much higher than the superconducting gap. We build tunneling conductance maps by plotting the normalized tunneling conductance for a given voltage, for example at zero bias. We also show below maps of the value of the superconducting gap and of the voltage where we observe a finite tunneling conductance (onset voltage). To obtain a map of the superconducting gap we plot the position in voltage of the quasiparticle peaks. We take the average between quasiparticle peak positions at positive and negative bias voltages. As we show below, at low enough temperatures, the quasiparticle peak position in the tunneling conductance coincides approximately with the superconducting gap value in the density of states. To map the onset voltage for a finite conductance we calculate at each point of the map the voltage at which the tunneling conductance is non-zero within experimental uncertainty when ramping from zero bias.

It is useful to remind that in STM we approach a metallic tip to the sample, until we observe vacuum tunneling at distance of a few \AA. The tunneling current is related to the work function's of tip and sample and to the overlap of their electronic wavefunctions, which can be related to the local density of states of tip and sample\,\cite{voigtlander2015scanning}. The actual value of the electronic density of states of the sample requires knowledge of the densities of states and their coupling, which is not available in STM\,\cite{PhysRevB.31.805}. Usually, good metals simply present a flat density of states a few meV around the Fermi level. This allows for an exact determination of the energy dependence of the density of states of samples close to the Fermi level. The most obvious case is possibly a superconductor, where a gap opens below T$_c$\,\cite{Suderow2014,RevModPhys.79.353}. The gap value, the shape of the quasiparticle peaks and the density of states inside the superconducting gap can be determined accurately, because the density of states of the tip (Au here) is completely flat in this energy range.

However, the absolute value of the electronic density remains unknown. Often, samples with strong disorder present an energy dependent normal density of states in the meV range, see for example Ref.\,\cite{Sacepe2011,Postolova2020,Carbillet2020}. The spatial fluctuations of the normal phase density of states can eventually be related to the spatial fluctuations of the superconducting gap\,\cite{Sacepe2011,Postolova2020,Carbillet2020}. As we discuss below, here we find essentially a flat density of states in the bias voltage range we have studied. Samples with larger degrees of disorder might present a reduction of screening and changes in the density of states around the Fermi level, as found in other systems\,\cite{Sacepe2011,Postolova2020,Carbillet2020}.

\section*{Results}
\begin{figure*}[ht!]
	\centering
		\includegraphics[width=0.9\linewidth]{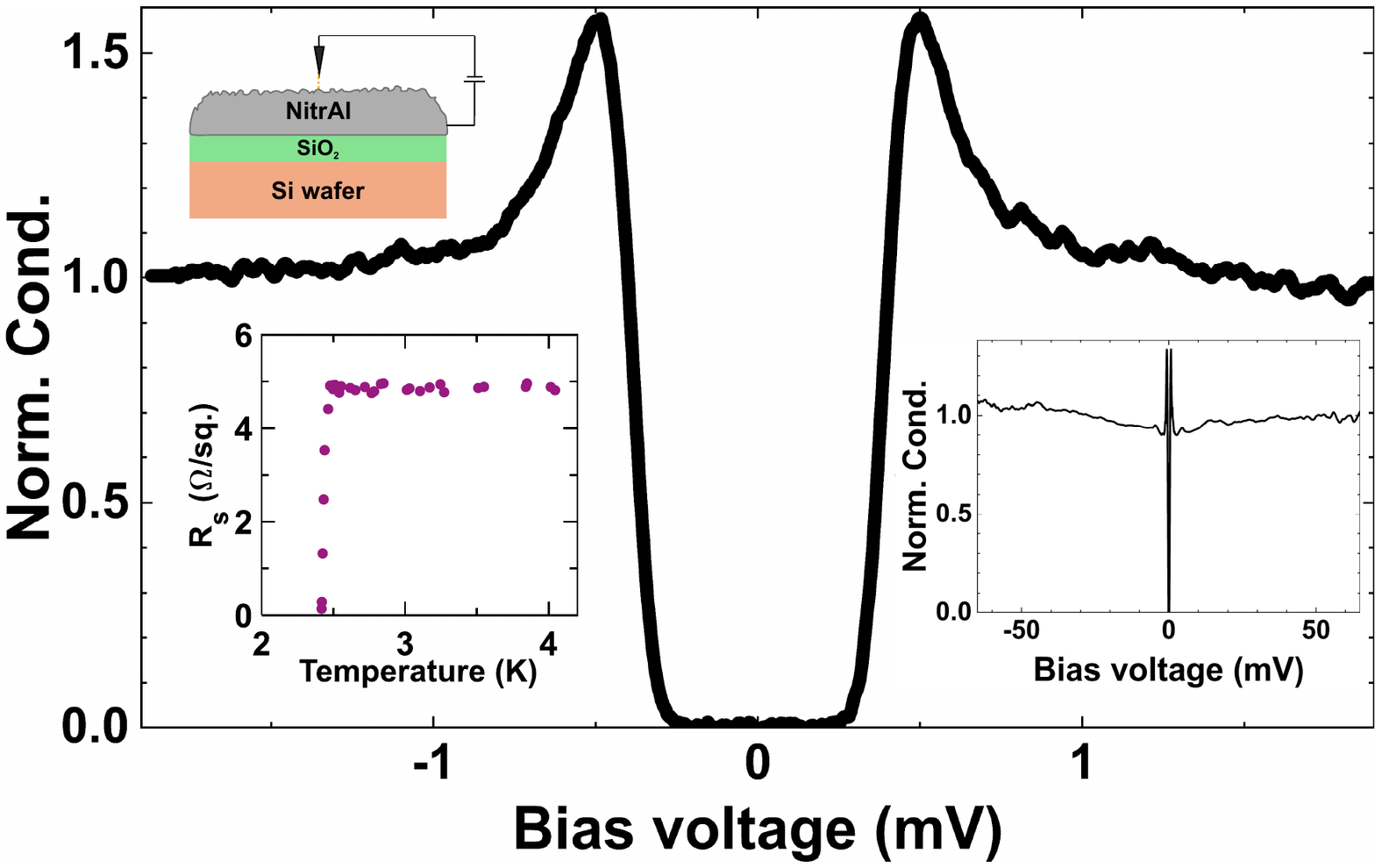} 
	\caption{Superconducting tunneling conductance on NitrAl thin film. Data were taken with a bias voltage of $V=2.5 ~\mathrm{mV}$, a current of $I=2~\mathrm{nA}$ and at a temperature $T=100~\mathrm{mK}$. Upper left inset shows a schematic representation of the tunneling experiment showing the NitrAl thin film in grey, the Si layers on the wafer in green and orange and the STM tip in black. Lower left inset shows the resistive transition of the measured film (data are sample ``G'' of Ref.\,\cite{TorrasColoma2024}). Right inset shows the tunneling conductance of NitrAl at a larger bias voltage range (taken starting at $V=70~\mathrm{mV}$, $I=3~\mathrm{nA}$ and at a temperature of $T=100~\mathrm{mK}$).}
		\label{fig:intro}
\end{figure*}

The superconducting NitrAl sample studied here shows typical metallic behavior with the opening of a superconducting gap at low voltages, as seen in Fig.\,\ref{fig:intro}. The metallic character is visualized by a nearly flat density of states for bias voltages up to 70$~\mathrm{mV}$ (see right inset in Fig.\,\ref{fig:intro}) and suggests that electrons remain mostly delocalized, as usual in metals\,\cite{Fernandez2021}.

\begin{figure*}[ht!]
	\centering
		\includegraphics[width=0.7\linewidth]{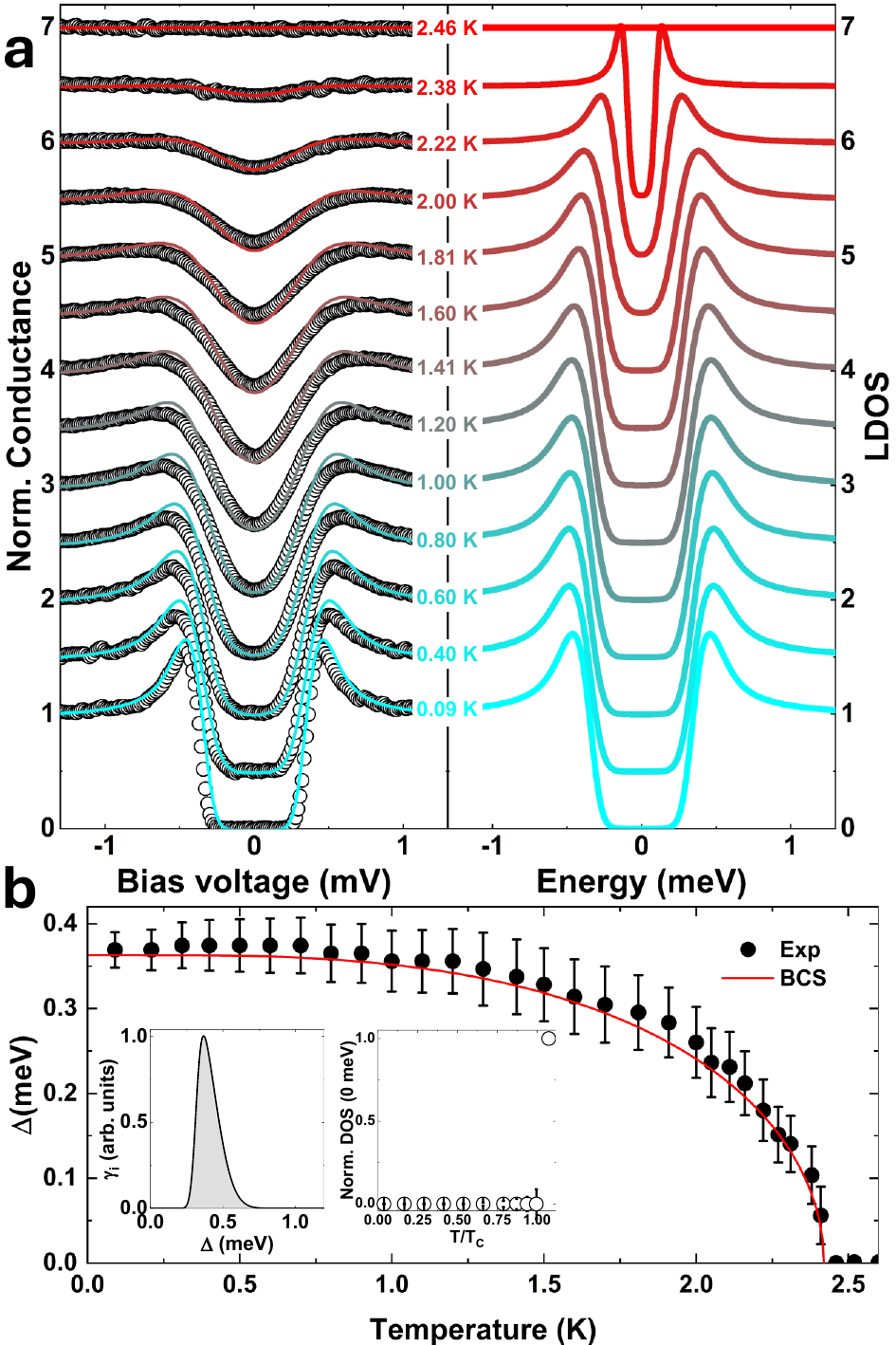} 
	\caption{Tunneling conductance of NitrAl film versus temperature. \textbf{a} Left panel shows the tunneling conductance vs bias voltage as a function of temperature (dots) and calculated conductance curves obtained after convoluting the DOS shown on the right panel with the derivative of the Fermi function at each temperature, as described in the text. Curves are shifted vertically for clarity. \textbf{b} Temperature dependence of the superconducting gap, obtained as described in the text. The solid line is the BCS temperature dependence. Left inset shows the skewed Gaussian distribution $\gamma_i(\Delta_i)$ with a peak at $\Delta_0=0.37~\mathrm{meV}$ used for the $T=100~\mathrm{mK}$ fit. Right inset shows the normalized DOS at zero energy. We find the error bars by performing reasonable fits to tunneling conductance curves with a finite density of states and using the standard deviation as the error.}
		\label{fig:gapvst}
\end{figure*}

In Fig.\,\ref{fig:gapvst}(a) we show the temperature dependence of the tunneling conductance at a representative position over the surface of the sample. We find $T_{\mathrm{c}}\sim2.4~\mathrm{K}$, consistent with the value reported from resistivity measurements ($T_{\mathrm{c}} = 2.43 ~\mathrm{K}$) \,\cite{TorrasColoma2024}.

To estimate the superconducting density of states, we consider that the tunneling conductance is given by $\sigma(V)=\int N(E)\frac{\partial f(E-eV)}{\partial E}dE$, where N(E) is the density of states, $E$ is the energy and $f(E)$ the Fermi function. Following Refs.\,\cite{PhysRevLett.101.166407, Herrera2023, Pablo2025, moreno2025}, we obtain the density of states $N(E)$ as a function of temperature by de-convoluting $\frac{\partial f(E-eV)}{\partial E}$ from $\sigma(V)$. The result is shown in Fig.\,\ref{fig:gapvst}(a). $N(E)$ deviates slightly from the usual s-wave density BCS expression $N(E)=Re(\frac{E}{\sqrt{\Delta^2-E^2}})$.

To address this deviation with more detail, let us first remark that we find that $N(E)$ remains zero for $E<0.25 \mathrm{meV}$ in the measured sample of NitrAl. To better understand this behavior, we assume a distribution of values of the superconducting gap, $\Delta_i$, and obtain $N(E) \propto \sum_i \gamma_i (\Delta_i)$Re$\left( E/\sqrt{E^2-\Delta^2} \right)$. We find that $\gamma_i (\Delta_i)$ has a Gaussian shape, skewed for low energies. Importantly, $\Delta_i>0.25$ meV, as shown in the lower left inset of Fig.\,\ref{fig:gapvst}(b), and there are no in-gap states below this value. Often, a lifetime broadened density of states is used to address tunneling spectroscopy of superconductors\,\cite{PhysRevLett.41.1509,PhysRevB.94.144508}. Such an approach does not apply here, however, because it induces a finite density of states at the Fermi level, and a finite $\sigma(V=0)$, not observed in this experiment.

On the other hand, the gap distribution $\gamma_i (\Delta_i)$ is peaked at $\Delta_0=0.37~\mathrm{meV}$. The shape of the gap distribution remains similar for all measured temperatures and has a width of $\pm 0.1~\mathrm{meV}$. The gap distribution is about twice larger than the one found in pure Pb or Al\,\cite{Fernandez2021, Fran2021}. $\Delta_0$ is very close to the expected BCS gap value $\Delta=1.76 k_B T_{\mathrm{c}}=0.36~\mathrm{meV}$, $T_{\mathrm{c}}=2.43~\mathrm{K}$. Importantly, as in Pb or Al \cite{Fernandez2021, Fran2021}, the density of states at zero energy remains zero until very close to T$_c$ (inset of Fig.\,\ref{fig:gapvst}(b)). In Fig.\,\ref{fig:gapvst}(b), we show the temperature evolution of $\Delta_0$ and compare it with the BCS prediction for $\Delta (T)$ obtained by solving self-consistently the gap equation. We observe a close agreement.

\begin{figure*}[ht!]
	\centering
		\includegraphics[width=0.9\linewidth]{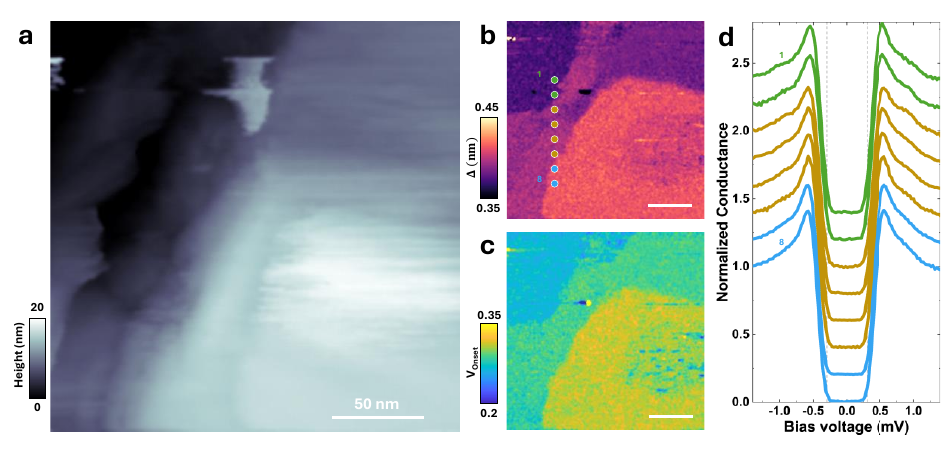} 
	\caption{Spatial dependence of the superconducting density of states of NitrAl at zero magnetic field and at $100~\mathrm{mK}$. \textbf{a} STM topography on NitrAl thin films. \textbf{b (c)} Superconducting gap width map (Onset voltage for finite conductance map) corresponding to the area shown in \textbf{a}. \textbf{d} Tunneling conductance curves acquired on the colored dots marked in \textbf{b}. We mark the first and last curves with numbers ''1" and ''8". We mark the position of $0.3~\mathrm{mV}$ with grey dashed vertical lines.}
		\label{fig:surface}
\end{figure*}

In Fig.\,\ref{fig:surface}(a) we show a topography of the NitrAl surface. The surface consists of separated areas with height differences of a few nm. The overall height variations amount up to 20 nm. Each of these areas is smooth at the nm level. We find a fully developed s-wave superconducting gap, shown in  Figs.\,\ref{fig:surface}d over the whole surface.

The tunneling conductance presents small changes as a function of the position, which can be correlated to topographic features. We highlight these differences in Fig.\,\ref{fig:surface}(b-d). In Fig.\,\ref{fig:surface}(b) we map the superconducting gap width in the same region as the topography (Fig.\,\ref{fig:surface} (a)). Likewise, in Fig.\,\ref{fig:surface}(c) we map the bias voltage for the onset of a finite conductance (i.e. tunneling conductance larger than 10\% of its value at about 1mV). In Fig. \ref{fig:surface} (d) we show tunneling conductance curves for several positions. Altogether, both the superconducting energy gap and the voltage corresponding to the onset of a finite $N(E)$ are spatially dependent, and vary up to $10\%$ across different probed regions.

\begin{figure*}[ht!]
	\centering
		\includegraphics[width=0.9\linewidth]{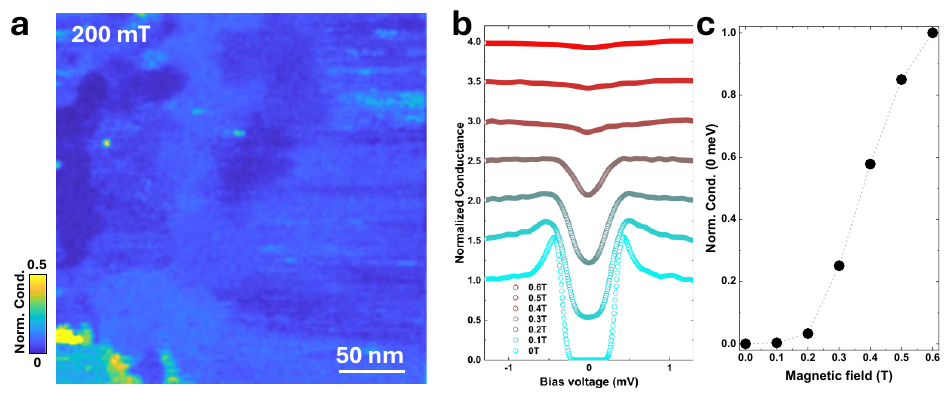} 
	\caption{Spatial dependence of the superconducting density of states of NitrAl at a finite magnetic field and at $100~\mathrm{mK}$. \textbf{a} Tunneling conductance map obtained at zero bias, $200~\mathrm{mT}$, and $100~\mathrm{mK}$. \textbf{b} Tunneling conductance curves as a function of applied magnetic field, taken on the areas of \textbf{a} with blue color. \textbf{c} Tunneling conductance value at zero voltage as a function of applied magnetic field. Values obtained from curves in \textbf{b}. We find that superconductivity vanishes altogether at about $\sim 500~\mathrm{mT}$ in the areas we have studied.}
		\label{fig:field}
\end{figure*}

Applying an external magnetic field perpendicular to the surface, the zero bias tunneling conductance $\sigma(V=0~\mathrm{mV})$ remains zero until about 0.1 T (Fig.\,\ref{fig:field}). Superconducting behavior is observed up to a field of about $0.6~\mathrm{T}$. In Fig.\,\ref{fig:field}(a) we show a zero-bias tunneling conductance map with a magnetic field of $200~\mathrm{mT}$. It is unclear why we do not observe vortices. We note that this often occurs in measuring thin films and remains difficult to explain\,\cite{Postolova2020,PhysRevB.88.014503,PhysRevB.103.214512,PhysRevB.100.214518}. We also note that under magnetic fields there are considerable spatial inhomogeneities of the superconducting density of states, up to 50\% of the normalized tunneling conductance, as shown by the color scale in Fig.\,\ref{fig:field}(a), suggesting an inhomogeneous upper critical field.

\section*{Discussion}
We have shown that superconductivity with $T_{\mathrm{c}}=2.4~\mathrm{K}$, and a sample surface smooth on the level of the nm, separated in areas with height difference of a few nm, are found for a NitrAl sample featuring a nearly temperature-independent resistivity. The surface is somewhat different from the one found in GrAl films, which displays grains with a circular shape a few nm in size, or the smooth surfaces found on the nm scale in TiN, NbN or other thin films\,\cite{Yang2020,kamlapure2010,Szabo2016,Kuzmiak2023,Zemlika2015,Yang2020,Bagwe2024,PhysRevB.102.060501,Postolova2020,Sacepe2011,Bagwe2024}. In NitrAl thin films it has been found that already low N doping reduces gran size and leads to smooth surfaces\,\cite{Lee2024,Lee2025}. Our observations suggest that grains are electronically very well connected with each other and lead to a spatially-smooth density of states.

Furthermore, we also show that the gap width changes only very weakly ($<$10~\%) between different positions on the surface of the sample. This suggests that either the composition of the film is very homogeneous, or there exist grains whose electronic coupling is so strong that spatial variations in the density of states remain un-noticed. By contrast, observations in thin films often display a finite superconducting subgap density of states at the Fermi level, particularly when approaching the superconductor-to-insulator transition\,\cite{kamlapure2010,Szabo2016,Kuzmiak2023,Zemlika2015,Yang2020,Bagwe2024,PhysRevB.102.060501}. Even nitride films far from exhibiting the superconductor-to-insulator transition, as TiN or NbN, often present local variations in the DOS which amount to more than $20~\%$\,\cite{Postolova2020,Sacepe2011,Bagwe2024}. Given the experimental uncertainty in the determination of the tunneling conductance inside the superconducting gap, we find that the subgap density of states is not larger than 2\% of the density of states above the superconducting gap. As the tunneling resistance for voltages above the gap is of about $1M\Omega$ (corresponding to a single atom junction), the tunneling resistance inside the gap is of $50 M\Omega$.

The interface between thin film superconductors and the substrate has been shown to possess pair-breaking effects on the superconductor\,\cite{Kuzmiak2022}. For example, it was shown that the zero bias conductance, critical temperature and the gap size considerably change depending on the substrate for thin films of MoC\,\cite{Haskova2018}. A more detailed study of NitrAl using different substrates would be required to elucidate this behavior.

We also find broad quasiparticle peaks. These have been reported before in disordered superconductors, such as TiN or NbN\,\cite{Gantmakher2010,Sacepe2011,chand2012,Burmistrov2016,Gurevich2017,Lotnyk2017,Postolova2020,Stosiek2021} and have been related to the suppression of phase coherence of the superconducting condensate accompanied by a reduction of the DOS and the superfluid density\,\cite{tinkham,Mondal2011,Seibold2012}. A reduced superfluid density can be associated to an enhanced kinetic inductance in disordered superconducting thin films\,\cite{tinkham,Masluk2012,Bell2012}. Here we have not observed granular electronic behavior previously observed on GrAl\,\cite{Yang2020} or on nitrides\,\cite{Sacepe2008}, neither Coulomb correlations effects on the low energy part of the density of states\,\cite{Butko2000,Carbillet2020}. However, the observed broadened quasiparticle peaks, with a width which exceeds the spatial changes in the gap size, could indicate enhanced Coulomb correlations. This suggests that our sample could also have a reduced superfluid density and eventually a high kinetic inductance. These properties are desirable for superinductors typically used in flux and fluxonium qubit applications\cite{Orlando1999,Manucharyan2009,Pop2014,Grunhaupt2019,Rieger2023}. 

\section*{Conclusions}
We have characterized the superconducting density of states of a NitrAl thin film down to 100$~\mathrm{mK}$ using a very low temperature STM. We find a superconducting density of states with well developed quasiparticle peaks whose position in energy agrees with BCS theory. The gap edge is well defined but it is located well below the quasiparticle peak and changes as a function of the position. The origin of the spatial inhomogeneities is not fully clear and could be related to morphological, structural and compositional differences typical in polycrystal-derived thin films. Optimization of thin film-substrate interface processing and deposition conditions, together with systematically studying the full range of NitrAl thin films could provide valuable insights for improving or tuning spatial inhomogeneities. Nevertheless, the large region in energy with a zero density of states suggests that NitrAl can be an excellent candidate to explore the fabrication of highly-coherent superconducting qubit devices.

\section*{Acknowledgements}
We acknowledge support by the Spanish Research State Agency (PID2023-150148OB-I00, TED2021-130546B\-I00, PDC2021-121086-I00 and CEX2023-001316-M, PID2020-114071RB-I00, RYC2019-028482- I, PCI2019-111838-2, PID2021-122140NB-C31, PID2021-122140NB-C32, PCI2024- 153468, CEX2023-001397-M) and the Comunidad de Madrid through projects TEC-2024/TEC-380 “Mag4TIC” and PIPF-2023/TEC-30683. We acknowledge the “QUASURF” project [SI4/PJI/2024-00199] funded by the Comunidad de Madrid through the direct grant agreement for the promotion and development of research and technology transfer at the Universidad Autónoma de Madrid. We have benefited from collaborations through EU program Cost CA21144 (Superqumap), from SEGAINVEX at UAM in the design and construction of STM and cryogenic equipment, and from the micronanofabs infrastructure. We also acknowledge the European Commission (QuantERA SiUCs and QRADES), and program ‘Doctorat Industrial’ of the Agency for Management of University and Research Grants (2024 DI 00004). IFAE is partially funded by the CERCA program of the Generalitat de Catalunya. This study was supported by MICIN with funding from European Union NextGenerationEU (PRTR-C17.I1) and by Generalitat de Catalunya.

\section*{Author contributions}
J.A.M. performed the measurements, wrote the manuscript and prepared the figures, with the help of P.G.T. and the supervision of E.H.V., I.G. and H.S.. A.T., G.R. and P.F. manufactured and characterized the sample, providing relevant information and supporting the data acquisition at all points. The study was devised by H.S., P.F. and G.R.. All authors contributed to writing and reviewed the manuscript.

\newpage

%\section*{Appendix}

%In Fig. \ref{fig:Pb} we show the temperature dependence of the superconducting gap for a sample of Pb, a well-known BCS superconductor. In Fig. \ref{fig:Pb} a) we show tunneling conductance curves as a function of temperature and the corresponding fit and DOS used in the convolution. We have used a single-valued superconducting gap $\Delta_{Pb}=1.32$ meV rather than a distribution of gaps to fit tunneling conductance curves. This leads to sharp quasiparticle peaks. In Fig. \ref{fig:Pb} b) we show the temperature dependence of the width of the superconducting gap. Since Pb is a strong-coupling superconductor, we find a superconducting gap $\Delta_{Pb}$ bigger than the prediction from BCS theory \cite{Rodrigo2004SS}, $\Delta_{Pb,BCS}=1.76 k_B T_{\mathrm{c}}=1.09$ meV.

%\begin{figure*}[ht!]
%	\centering
%		\includegraphics[width=0.7\linewidth]{Pbv1.pdf} 
%	\caption{\textbf{a} On left panel we show the tunneling conductance vs bias voltage as a function of temperature (dots) and calculated conductance curves obtained after convoluting the BCS DOS (shown on the right panel) with the derivative of the Fermi function at each temperature, as described in the text. Curves are shifted vertically for clarity. \textbf{b} Temperature dependence of the superconducting gap, obtained as described in the text. Red solid line is the BCS temperature dependence that best fits our experimental data. We use $1.2 \cdot \Delta=1.32$ meV with $T_{\mathrm{c}} \sim 7.3$ K. Inset shows the normalized DOS at zero energy.}
%		\label{fig:Pb}
%\end{figure*}
\bibliography{biblio}  

\end{document}